\pdfoutput=1
\documentclass[final]{elsart}

\usepackage{graphicx}
\usepackage{amsmath}
\usepackage{amssymb}
\usepackage{times}
\usepackage{verbatim}

\DeclareGraphicsExtensions{.pdf}
\usepackage{amssymb}

\begin{document}

\begin{frontmatter}



\title{Mutual Trust and Cooperation in the Evolutionary Hawks-Doves Game}


\author[label1]{Marco Tomassini},
\author[label1]{Enea Pestelacci},
\author[label2]{Leslie Luthi}

\address[label1]{Information Systems Department, HEC, University of Lausanne,Switzerland}
\address[label2]{Formerly at the Information Systems Department, HEC, University of Lausanne,Switzerland}

\begin{abstract}
Using a new dynamical network model of society in which pairwise interactions are weighted according to
mutual satisfaction, we show that cooperation is the norm in the
Hawks-Doves game when individuals are allowed to break ties with undesirable neighbors
and to make new acquaintances in their extended neighborhood. Moreover, cooperation is robust
with respect to rather strong strategy perturbations. 
We also discuss the empirical structure of the emerging networks,
and the reasons that allow cooperators to thrive in the population.
Given the metaphorical
importance of this game for social interaction, this is an encouraging positive result as
standard theory for large mixing populations prescribes that  a certain fraction of defectors
must always exist at equilibrium. 

\end{abstract}

\end{frontmatter}

%\maketitle

\section{Introduction and Previous Work}
\label{intro}

Game Theory~\cite{vega-redondo-03} is the study of how social or economical agents take decisions
in situations of conflict. Some  games such as the celebrated Prisoner's Dilemma have a high metaphorical value for society in spite of their simplicity and abstractness. Hawks-Doves, also known as Chicken, is one such socially significant game.
Hawks-Doves is a two-person, symmetric game with the generic payoff bi-matrix of Table~\ref{payoffs}.
\begin{table}[hbt]
\vspace{-0.1cm}
\begin{center}
{\normalsize
\begin{tabular}{c|cc}
 & C & D\\
\hline

C & (R,R) & (S,T)\\
D & (T,S) & (P,P)
\end{tabular}
}
\end{center}  
\caption{Payoff matrix for a symmetric two person game.}   
\label{payoffs}
\vspace{-0.1cm}
\end{table}
\noindent In this matrix, D stands for the  defecting strategy ``hawk'', and C stands for the
cooperating strategy ``dove''.
The ``row'' strategies correspond to player 1 and the ``column'' strategies to player 2.
An entry of the table such as (T,S) means that if player 1 chooses strategy D and
player 2 chooses strategy C, then the payoff or utility to player 1 is T, while the payoff
of player 2 is S.
Metaphorically, a hawkish behavior means a strategy of fighting, while a dove, when facing a confrontation, will always yield. 
R is the \textit{reward}
the two players receive if they
both cooperate, P is the \textit{punishment} for bilateral defection, and T  is the
\textit{temptation}, i.e. the payoff that a player receives if it defects, while the
other cooperates. In this case, the cooperator gets the \textit{sucker's payoff} S.
The game has a structure similar to that of  the \textit{Prisoner's Dilemma}~\cite{axe84}.
However, the ordering of payoffs for the Prisoner's
Dilemma is $T > R > P > S$ rendering defection the best rational individual choice, while
in the Hawks-Doves game studied here the ordering is  $T > R > S > P$ thus making
mutual defection, i.e. result (D,D),  the worst possible outcome. Note that in  game theory, as long as the
above orderings are respected, the actual numerical payoff values do not change the nature and  number
of equilibria~\cite{vega-redondo-03}.\\
In contrast to the Prisoner's Dilemma which has a unique Nash equilibrium that corresponds to
both players defecting, 
the strategy pairs (C,D) and (D,C) are both Nash equilibria of the Hawks-Doves game in pure strategies, and there is
a third equilibrium in mixed strategies in which strategy D is played
with probability $p$, and strategy C with probability $1-p$, where $0 < p < 1$ depends on the actual
payoff values. We recall that a Nash equilibrium is a combination of strategies (pure or mixed)
of the different players such that any unilateral deviation by any agent from this combination can only decrease her expected payoff~\cite{vega-redondo-03}.\\
As it is the case for the 
Prisoner's Dilemma (see for example~\cite{axe84,lindnor94a} for the iterated case, among a vast literature), Hawks-Doves, for all its simplicity, appears to
capture some important features of  social interactions. In this sense, it applies
in many situations in which ``parading'', ``retreating'', and ``escalating'' are common.
One striking example of a situation
that has been thought to lead to a Hawks-Doves dilemma is the Cuban missile crisis in 1962
~\cite{poundstone92}. Territorial threats at the border between nations are another case in point
as well as bullying in teenage gangs.
Other well known applications are found in the animal kingdom during ritualized fights
 ~\cite{maynard82}.\\
 In this article, we shall present our methods and results in the framework of
 \textit{evolutionary game theory}~\cite{hofb-sigm-book-98}. In  evolutionary game theory a
 very large mixing population of players is considered, and
 randomly chosen pairs of individuals play a sequence of one-shot two-person games. In the Hawks-Doves game, the theory
prescribes that the only \textit{Evolutionary
Stable Strategy} (ESS) of the population is the mixed strategy, giving rise, at equilibrium,
to a polymorphic population composed of hawks and doves in which the frequency of hawks equals  $p$, the probability with which strategy
hawk would be played in the NE  mixed strategy.\\
In the case of the Prisoner's Dilemma, one finds a unique ESS with all the individuals defecting.
However, Nowak and May~\cite{nowakmay92} showed
that cooperation in the population is sustainable under certain conditions,
provided that the network of the interactions between players has a lattice spatial structure. Killingback and
Doebeli~\cite{KD-96} extended the
spatial approach to the Hawks-Doves game and found that a planar lattice structure
with only nearest-neighbor interactions may favor cooperation, i.e. the fraction of doves in
the population is often higher than what is predicted by evolutionary game theory. In a more recent work however, Hauert and Doebeli~\cite{hauer-doeb-2004} were led to a different conclusion, namely that
spatial structure does not seem to favor cooperation in the Hawks-Doves game.\\
Further studies extended the structured population approach to other graph structures
representing small worlds (for an excellent review, see~\cite{Szabo-Fath-07}). Small-world networks are produced by randomly rewiring a few links in
an otherwise regular lattice such as a ring or a grid~\cite{watts-strogatz-98}. These ``shortcuts'', as they are called,
give rise to graphs that have short path lengths between any two nodes in the average as in random graphs, but in contrast to the latter,
also have a great deal of local structure as conventionally measured by the \textit{clustering
coefficient}\footnote{The clustering coefficient
${\cal C}_{i}$ of a node $i$ is defined as ${\cal C}_i=2E_i/k_i(k_i-1)$, where $E_i$ is the number of edges in  the
neighborhood of $i$. Thus ${\cal C}_i$ measures the amount of ``cliquishness'' of the
neighborhood of node $i$ and it characterizes the extent to which nodes adjacent to node $i$ are
connected to each other. The clustering coefficient of the graph is simply the average over all nodes:
${\cal C} = \frac{1}{N} \sum_{i=1}^{N}{ \cal C}_i$~\cite{newman-03}.}. These structures are much more typical of the networks that have been analyzed
in technology, society, and biology than regular lattices or random graphs~\cite{newman-03}.
 In~\cite{tom-luth-giac-06} it was found that cooperation in Hawks-Doves may be either enhanced or inhibited in small-world
networks depending on the gain-to-cost ratio $r={R}/{(R-P)}$, and on the strategy update rule using standard
 local evolutionary
dynamics with one-shot bilateral encounters. However, Watts--Strogatz
small-world networks, although more realistic than lattices or random graphs, are not good representations of typical social networks.
Santos and Pacheco~\cite{santos-pach-05} extended the study of the Hawks-Doves game to scale-free
networks, i.e.~to networks having a power-law distribution of the connectivity degree~\cite{newman-03}. They found that
cooperation is remarkably enhanced in them with respect to previously described population
structures through the existence of highly connected cooperator hubs. 
Scale-free networks are much closer than Watts--Strogatz ones to
the typical  socio-economic networks that have been investigated, but they are relatively uncommon
in their ``pure'' form due to finite cutoffs and other real-world effects (for example, 
see~\cite{newman-03,am-scala-etc-2000,newman-collab-01-1,jordano03}),
 with the notable exception of sexual contact networks~\cite{lil-et-al-01}. 
 Using real and
model static social networks, Luthi et al.~\cite{luthi-pest-tom-physa07} also found that cooperation is
enhanced in Hawks-Doves, although to a lesser degree than in the scale-free case,
thanks to the existence of tight clusters of cooperators that reinforce each other.\\
Static networks resulting from the analysis of actual social networks or good models of the latter are a good starting point;
 however, the static approach ignores 
fluctuations and non-equilibrium phenomena.  As a matter of fact, in many 
real networks nodes
may join the network forming new links, and old nodes may leave it as social actors
come and go. Furthermore, new links between agents already in the network may also form or be
dismissed. Often the speed of these network changes is comparable to that of the
agent's behavioral adaptation, thus making it necessary to study how they interact.
Examples of slowly-changing social networks are scientific collaborations, friendships, firm networks
among others. A static network appears to be a good approximation in these cases.
On the other hand, in our Internet times, there exist many social or pseudo-social
networks in which topology changes are faster. For example, e-mail networks~\cite{kossi-watts-06},
web-based networks for friendship and entertainment, such as Facebook, or professional purposes such as LinkedIn, and many others. Furthermore, 
as it is not socially credible that people will keep for a long time unsatisfying relationships,
addition and dismissal of partners are an extremely common phenomenon, also due to natural causes
such as moving, changing fields, or interests.
We note at this point that some previous work has focused on the possibility of allowing players to choose
or refuse social partners in game interactions~\cite{batali,sherratt}, which has been shown to potentially promote
cooperation. However, this work does not consider an explicit underlying
interaction network of agents, nor does it use the social link strengths as indicators of partner's suitability as
we do here.\\
In light of what has been said above, the motivation of the present work is to study the co-evolution of strategy  and
network structure and to investigate under which conditions cooperative  behavior may emerge and be stable in the Hawks-Doves game. 
A related goal is to study the topological structures of the emergent networks and their relationships with the strategic choices of the agents.
Some previous work has been done on evolutionary games on dynamic networks
~\cite{skyrms-pem-00,eguiluz-et-al-05,zimmer-pre-05,lut-giac-tom-al-06,santos-plos-06} almost all of them dealing with
the Prisoner's Dilemma. The only one briefly describing results for the Hawks-Doves
game is~\cite{santos-plos-06} but our model differs in several important respects and we obtain
new results on the structure of the cooperating clusters. The main novelty is the use
of pairwise interactions that are dynamically weighted according to mutual satisfaction.
The new contributions and the differences with
previous work will be described at the appropriate points in the article.
An early preliminary version of this study has been presented at the conference~\cite{pest-tom-hd-08}.\\
The paper is organized as follows. In the next section we present our coevolutionary model. This is
followed by an exhaustive numerical study of the game's parameter space. After that we present
our results on cooperation and we describe
and discuss the structure of the emerging networks. Finally we give our conclusions and suggestions
for possible future work.

\section{The Model and its Dynamics}
\label{mod-dyn}

The model is strictly local as no player uses information other than the one concerning the player itself and 
the players it is directly connected to. In particular, each agent knows its own current strategy and payoff.
Moreover, as the model is an evolutionary one, no rationality, in the sense of
game theory, is needed~\cite{vega-redondo-03}. Players just adapt their behavior such that they imitate more successful strategies in their environment  with higher probability. Furthermore, they are able to locally
assess the worthiness of an interaction and possibly dismiss a relationship that does not pay off enough.
The model has been introduced and fully explained in~\cite{pest-tom-lut-08}, where we study the Prisoner's Dilemma and
the Stag-Hunt games; it is reported here in some detail in order to make the
paper self-contained.

\subsection{Agent-Agent and Network Interaction Structure}
\label{net}
\begin{figure} [!ht]
\begin{center}
%	\vspace*{-0.2cm}
\includegraphics[width=4.5cm] {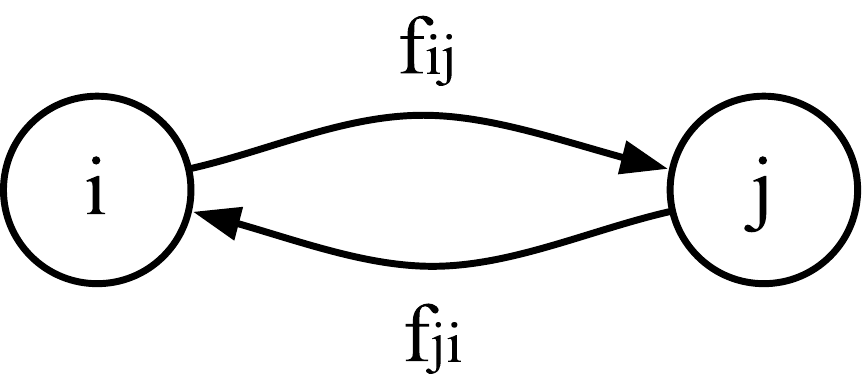} \protect \\
%\vspace{-0.2cm}	
\caption{Schematic representation of mutual trust between two agents through the strengths
of their links.\label{force}}
\end{center}
\end{figure}

The network of agents is represented by a directed graph $G(V,E)$, where the
 set of vertices $V$ represents the agents, while the set of oriented edges (or links) $E$ represents their unsymmetric  interactions. The
 population size $N$ is the cardinality of $V$. A neighbor  of an agent $i$ is any other agent $j$ such that there is a pair of oriented
 edges $\vec {ij}$ and $\vec {ji} \in E$.
The set of neighbors of $i$  is called  $V_i$. For network structure description purposes, we shall also use an unoriented version $G^{'}$
of $G$ having exactly the same set of vertices $V$ but only a single unoriented edge $ij$ between any pair of connected vertices
$i$ and $j$ of $G$. For $G{'}$ we shall define the degree  $k_i$ of vertex $i \in V$ as the number of neighbors of $i$. The average
degree of the network $G^{'}$ will be called $\bar k$.\\
A pair of directed links between vertices $i$ and $j$ in $G$ is schematically depicted in Fig.~\ref{force}. Each
link has a weight or ``force'' $f_{ij}$ (respectively $f_{ji}$). This weight, say $f_{ij}$,  represents in an indirect way the 
``trust'' player $i$ attributes to player $j$. This weight may take any value in $[0,1]$ and its
variation is dictated by the payoff earned by $i$ in each encounter with $j$, as explained below.

The idea behind the introduction of the forces $f_{ij}$ is loosely inspired by the potentiation/depotentiation of
connections between neurons in neural networks, an effect known as the \textit{Hebb rule} \cite{hebb}. In our context, it can be seen
as a kind of  ``memory'' of previous encounters. However, it must be distinguished from the memory used in
 iterated games, in which
players ``remember'' a certain number of previous moves and can thus conform their future strategy on the
analysis of those past encounters~\cite{vega-redondo-03}. Our interactions are strictly one-shot, i.e.~players ``forget'' the results of
previous rounds and cannot recognize previous partners and their possible playing patterns. However, a certain amount of past history is implicitly
contained in the numbers $f_{ij}$ and this information may be used by an agent when it will come to decide
whether or not an interaction should be dismissed (see below). \\
We also define a quantity $s_i$ called \textit{satisfaction} of an agent $i$ which is the sum of all the weights
of the links between $i$ and its neighbors $V_i$ divided by the total number of links $k_i$:

$$ s_i = \frac{\sum_{j \in V_i} f_{ij} } {k_i}. $$

\noindent We clearly have $0 \le s_i \le 1$. Note that the term satisfaction is sometimes used in game-theoretical
work to mean the amount of utility gained by a given player. Instead, here satisfaction is related to the average willingness
of a player to maintain the current relationships in the player's neighborhood.

\subsection{Initialization}
\label{init}

The network is of constant size  $N=1000$; this allows a simpler yet significant model of network
dynamics in which social links may be broken and formed but agents do not disappear and new
agents may not join the network. The initial graph is generated randomly with
a mean degree  $\bar k=10$ which is  of the order of those actually found in many social networks such
as collaboration, association, or friendship networks in which relations are generally rather long-lived and there is a cost to maintain a large number; see, for
instance,~\cite{newman-collab-01-1,newman-03,moody-04,TLGL-GPEM-07}.
Players are distributed uniformly at random over the graph vertices with 50\% cooperators. Forces of links
between any pair of neighboring players are initialized at $0.5$.\\
We use a parameter $q$ which is akin to a
``temperature'' or noise level; $q$ is a real number in $[0,1]$ and
it represents the frequency with which an agent wishes to dismiss a link with one of its neighbors. The higher
$q$, the faster the link reorganization in the network. This parameter has been first introduced in~\cite{zimmer-pre-05} 
and it controls the speed at which topological changes occur in the network, i.e. the time scale of the strategy-topology
co-evolution. It is
an important consideration, as social networks may structurally evolve at widely different speeds, depending
on the kind of interaction between agents. For example, e-mail networks change their structure at a faster pace 
than, say, scientific collaboration networks.

\subsection{Strategy and Link Dynamics}
\label{strat-link-dyn}

Here we describe in detail how individual strategies, links, and link weights are updated. The node update sequence is chosen at
random with replacement as in many previous works~\cite{hubglance93,hauer-doeb-2004,lut-giac-tom-al-06}. Once a given
node $i$ of $G$ is chosen to be activated, it goes through the following steps:

\begin{itemize}
\item if the degree of agent $i$, $k_i = 0$ then
 player $i$ is an isolated node. In this case a link with strength $0.5$ is created
from $i$ to a player $j$ chosen uniformly at random among the other $N-1$ players in the network. 
\item otherwise,
\begin{itemize}
\item either  agent $i$ updates its strategy according to a local \textit{replicator dynamics} rule with probability $1-q$ or, with probability $q$, agent $i$ may delete a link with a given neighbor $j$ and creates a new $0.5$ force link with another node $k$ ;

\item the forces between $i$ and its neighbors $V_i$ are updated 

\end{itemize}
\end{itemize}

Let us now describe each step in more detail.

\subsection{Strategy Evolution}

We use a local version of replicator dynamics (RD) for regular graphs~\cite{hauer-doeb-2004} but modified as described
in~\cite{luthi-tomas-pest-09} to take into account the fact that the number of neighbors in a degree-inhomogeneous
network can be different for different agents. Indeed, it has been analytically shown that using straight accumulated payoff
in degree-inhomogeneous networks leads to a loss of invariance with respect to affine transformations of the payoff
matrix under RD~\cite{luthi-tomas-pest-09}.
The local dynamics of a player $i$ only depends on its own strategy and on the strategies of the $k_i$ players in its neighborhood 
$V_i \in G^{'}$.
Let us call $\pi_{ij}$ the payoff player $i$ receives when interacting with neighbor $j$. This payoff is defined as
$$
\pi_{ij} =  \sigma_i(t)\; M\; \sigma_{j}^T(t),
$$
\noindent where $M$ is the payoff matrix of the game and $\sigma_i(t)$ and $\sigma_j(t)$ are the strategies played by $i$ and $j$ at time $t$.
The quantity
$$
 \widehat{\Pi}_i(t) = \sum _{j \in V_i}\pi_{ij}(t)
$$
\noindent is the weighted accumulated payoff defined in~\cite{luthi-tomas-pest-09} collected by player $i$ at time step $t$.
The rule according to which agents update their strategies is the conventional RD in which strategies that do
better than the average increase their share in the population, while those that fare worse than average
decrease.
To update the strategy of player $i$, another player $j$ is drawn at random from the neighborhood $V_i$.
It is assumed that the probability of switching strategy is a function $\phi$ of the payoff difference;
$\phi$ is
required to be monotonic increasing; here it has been taken linear~\cite{hofb-sigm-book-98}. Strategy $\sigma_i$ is replaced by $\sigma_j$ with probability

$$
 p_i = \phi(\widehat{ \Pi}_j - \widehat{\Pi}_i),
$$

where

%\begin{eqnarray}
$$
\phi(\widehat{\Pi}_j -\widehat{ \Pi}_i)  =
\begin{cases} \dfrac{\widehat{\Pi}_j - \widehat{\Pi}_i}{\widehat{\Pi}_{j,\textrm{max}} - \widehat{\Pi}_{i,\textrm{min}}} & \textrm{if $\widehat{\Pi}_j - 
\widehat{\Pi}_i > 0$}\\\\
0 & \textrm{otherwise.}
\end{cases}
%\label{repl_dyn_eq2}
%\end{eqnarray}
$$
In the last expression, $\widehat{\Pi}_{x,\textrm{max}}$ (resp.\ $\widehat{\Pi}_{x,\textrm{min}}$) is the maximum (resp.\ minimum) payoff a player $x$ can get (see ref.~\cite{luthi-tomas-pest-09} for more details).
 
The major differences with standard RD is that two-person encounters between players are only possible among neighbors, instead of being drawn from the whole population, and the latter is of finite size in our case.
Other commonly used strategy update rules include imitating the best in the 
neighborhood~\cite{nowakmay92,zimmer-pre-05}, or replicating in proportion to the payoff ~\cite{hauer-doeb-2004,tom-luth-giac-06}.

\subsection{Link Evolution}

\begin{figure} [!ht]
\begin{center}
%	\vspace*{-0.2cm}
\includegraphics[width=6cm] {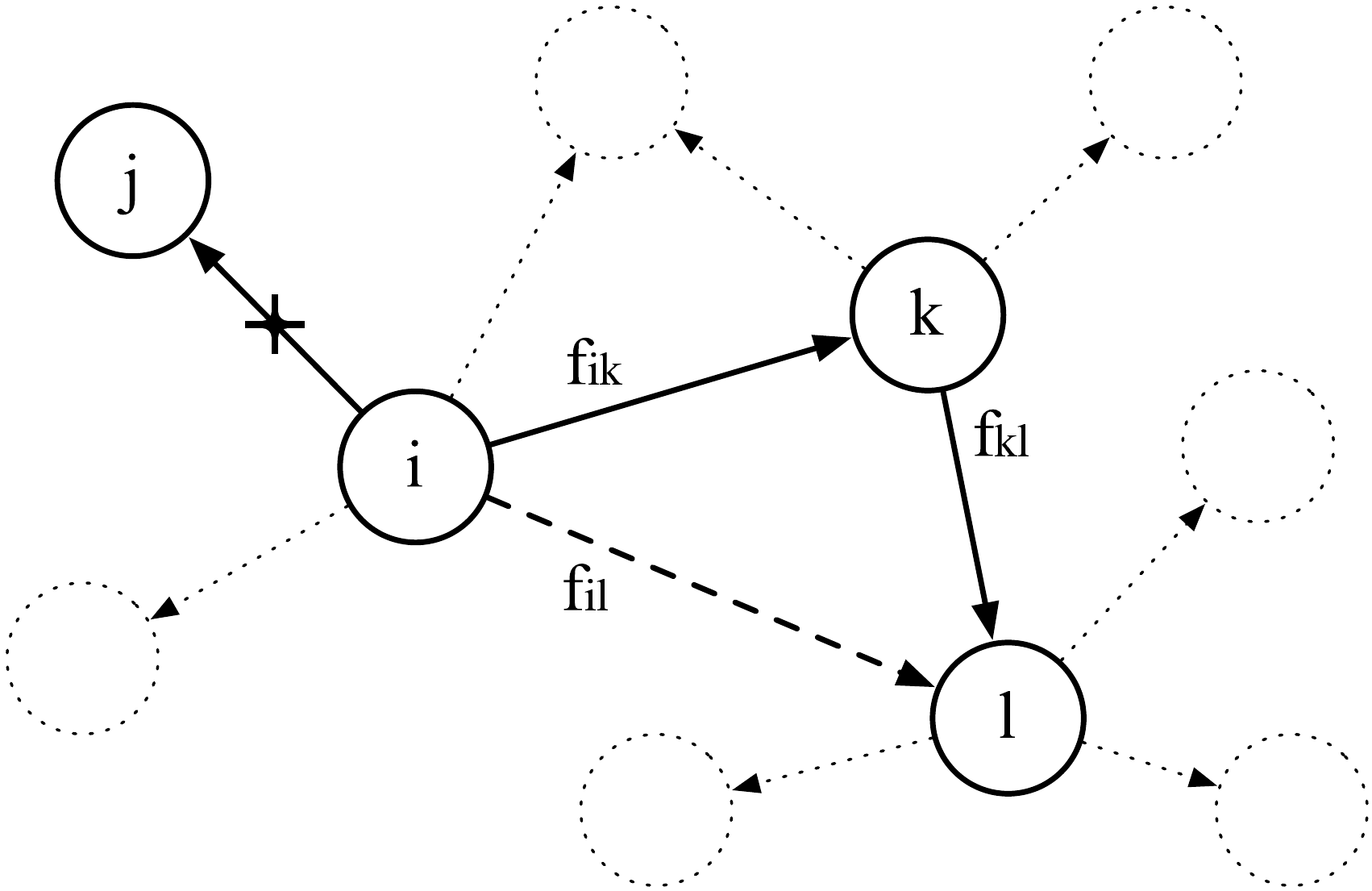} \protect \\
%	\vspace*{-0.2cm}
\caption{Illustration of the rewiring of link $\{ij\}$ to $\{il\}$. Agent $k$ is
chosen to introduce player $l$ to $i$ (see text). Only outgoing links are shown for clarity. \label{rewire}}
\end{center}
\end{figure}

The active agent $i$, which has $k_i \ne 0$ neighbors will, with probability $q$, attempt to dismiss an interaction with one of its neighbors in the following way. In the description we focus on the outgoing links from $i$ in $G$, the incoming links play a subsidiary role.
Player $i$ first  looks at its satisfaction $s_i$. The higher $s_i$, the more satisfied the player, since a high satisfaction
is a consequence of successful strategic interactions
with the neighbors. Thus, the natural tendency is to try to dismiss a link when $s_i$ is low. This is simulated by drawing a uniform pseudo-random number $r \in [0,1]$ and breaking a link when $r \ge s_i$.
Assuming that the decision is taken to cut a link, which one, among the possible $k_i$, should be chosen?
Our solution is based on the strength of the relevant links. First a neighbor $j$ is chosen with probability proportional to $1-f_{ij}$, i.e. the stronger the link, the less likely it is that it will be selected. This intuitively corresponds to $i$'s observation that it is preferable to dismiss an interaction with a neighbor $j$ that has contributed little to $i$'s payoff over several rounds of play. However, dismissing a link is not free: $j$ may
``object'' to the decision. The intuitive idea is that, in real social situations, it is seldom possible to take unilateral
decisions: often there is a cost associated, and we represent this hidden cost by a probability 
$1 - (f_{ij} + f_{ji})/2$
with which $j$ may refuse to be cut away. In other words, the link is less likely to be deleted if $j$ appreciates $i$, i.e. when $f_{ji}$ is high.\\
Assuming that the $\vec {ij}$ and $\vec {ji}$ links are finally cut, how is a new interaction to be formed? 
The solution adopted here is inspired by the observation that, in social settings, links are
usually created more easily between people who have a mutual acquaintance than those who do not.
First, a neighbor $k$ is chosen in $V_i \setminus \{j\}$ with probability proportional to $f_{ik}$, thus favoring neighbors $i$ trusts.
Next, $k$ in turn chooses player $l$ in his neighborhood $V_k$ using the same principle, i.e. with probability proportional to $f_{kl}$. If $i$ and $l$ are not connected,
two links $\vec {il}$ and $\vec {li}$ are created, otherwise the process is repeated in $V_l$. Again, if the selected node, say $m$, is not connected to $i$, an interaction between $i$ and $m$ is established by creating two new links $\vec {im}$ and $\vec {mi}$. If this also fails,  new links between $i$ and a randomly chosen node are created. 
In all cases the new links are initialized with a strength of $0.5$ in each direction.
This rewiring process is schematically depicted in Fig.~\ref{rewire} for the case in which a link can be successfully established between players $i$ and $l$ thanks to their mutual acquaintance $k$.

At this point, we would like to stress several important differences with previous work in which links can be
dismissed and rewired in a constant-size network in evolutionary games. First of all, in all these works the
interaction graph is undirected with a single link between any pair of agents. In~\cite{zimmer-pre-05}, only links between 
defectors are allowed to be cut unilaterally and the study is restricted to the Prisoner's Dilemma. Instead,
in our case, any interaction has a finite probability to be abandoned, even a profitable one between cooperators if it is
recent, although
links that are more stable, i.e. have high strengths, are less likely to be rewired. This smoother situation is made possible thanks to
our bilateral view of a link. It also allows for a moderate amount of ``noise'', which could
 reflect to some extent the uncertainties in the system. The present link rewiring process is also different from
 the one adopted in~\cite{santos-plos-06}, where the Fermi function is used to decide whether to cut a link or not and
 also from their new version of it which has appeared in~\cite{vanSegbroeck}.
 Finally, in~\cite{lut-giac-tom-al-06} links are cut according to a threshold decision rule and are rewired randomly anywhere in the network.

\subsection{Updating the Link Strengths}

Once the chosen agents have gone through their strategy or link update steps, the strengths of the
links are updated accordingly in the following way:

$$ f_{ij}(t+1) = f_{ij}(t) + \frac {\pi_{ij} - \bar\pi_{ij}}  {k_i(\pi_{max} - \pi_{min}) },    $$

\noindent where $\pi_{ij}$ is the payoff of $i$ when interacting with $j$, $\bar\pi_{ij}$ is the payoff earned by
$i$ playing with $j$, if $j$ were to play his other strategy, and $\pi_{max}$ ($\pi_{min}$) is the maximal (minimal) possible payoff obtainable in a single interaction. 
If $f_{ij}(t+1)$ falls outside the $[0,1]$ interval then it is reset to $0$ if it is negative, and to $1$ if it is larger than $1$.
This update is performed in both
directions, i.e. both $f_{ij}$ and $f_{ji}$ are updated $\forall j \in V_i$ because
both $i$ and $j$ get a payoff out of their encounter.

\section{Numerical Simulations and Discussion}
\label{simul}

\subsection{Simulation Parameters}
\label{sim-par}

We simulated the Hawks-Doves game with the dynamics described above exploring the game space by limiting our study to the variation
of only two game parameters. 
We set $R=1$ and $P=0$ and the two parameters are $1 \leq T \leq 2$ and $0 \leq S \leq 1$.
Setting $R=1$ and $P=0$ determines the range of $S$ (since  $T>R>S>P$)
and gives an upper bound of 2 for $T$, due to the $2R > T+S$ constraint, which ensures that
mutual cooperation is preferred over an equal probability of unilateral cooperation and defection.
Note however, that the only valid value pairs of $(T,S)$ are those that satisfy the latter constraint.

\begin{figure*} [!ht]
\begin{center}
%	\vspace*{-0.2cm}
\includegraphics[width=14.5cm] {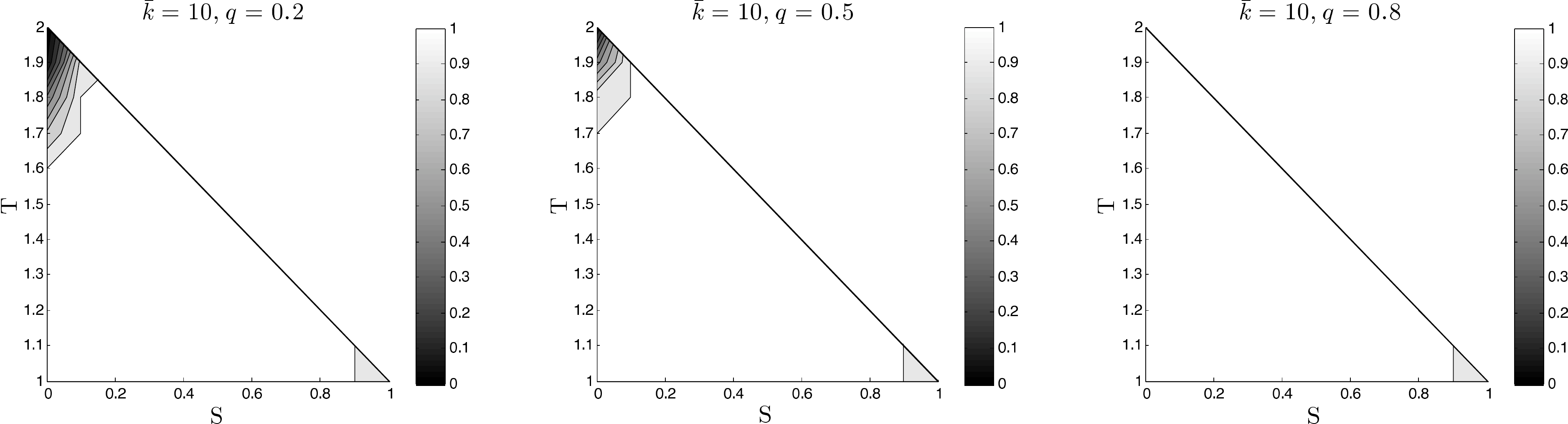} \protect \\	
%	\vspace*{-0.2cm}
\caption{Average cooperation values for the Hawks-Doves game for three values of $q$ at steady-state.
Results are the average of $50$ runs.
\label{hd_coop}}
\end{center}
\end{figure*}

We simulated networks of size $N=1000$, randomly generated with an average degree $\bar k=10$ and randomly initialized with 50\% cooperators and 50\% defectors.
In all cases,  parameters $T$ and $S$ are varied between their two bounds in steps of 0.1.
For each set of values, we carry out 50 runs of at most 10000 steps each, using
a fresh graph realization in each run. Each step consists in the update of a full population.
A run is stopped when all agents are using the same strategy, in order to 
be able to measure statistics for the population and for the structural parameters of the graphs. After an initial transient period, the system is considered to have reached
a pseudo-equilibrium strategy state when the strategy of the agents (C or D) does not change over 150 further
time steps, which means $15 \times 10^4$ individual updates. It is worth mentioning that equilibrium is
always attained well before the allowed $10000$ time steps, in most cases, less than $1'000$ steps are enough.
We speak of pseudo-equilibria or steady states and not
of true evolutionary equilibria because there is no analog of equilibrium conditions in the dynamical systems sense.\\
To check whether scalability is an issue for the system, we have run several simulations with larger graphs namely,
$N=3000$ and $N=10000$. The overall result is that, although the simulations take a little longer and transient times
are also slightly longer, at quasi-equilibrium all the measures explored in the next sections follow the same trend
and the dynamics give rise to comparable topologies and strategy relative abundance.

\subsection{Emergence of Cooperation}
\label{coop}

Cooperation results  in contour plot form are shown in Fig.~\ref{hd_coop}. We remark that,
as observed in other structured populations, cooperation is achieved in almost  the whole configuration
space. Thus, the added degree of freedom represented by the possibility of refusing a partner and choosing
a new one does indeed help to find player's arrangements that help cooperation.  When considering the dependence on the 
parameter $q$, one sees in Fig.~\ref{hd_coop} that the higher $q$, the higher the cooperation level, although the differences are small, since full cooperation prevails already at $q=0.2$. This is a somewhat expected result, since
being able to break ties more often clearly gives cooperators  more possibilities for finding and keeping
fellow cooperators to interact with. The results reported in the figures are for populations starting with $50\%$
cooperators randomly distributed. We have also tried other proportions with less cooperators, starting at $30\%$.
The results, not reported here for reasons of space, are very similar, the only difference being that
it takes more simulation time to reach the final quasi-stable state. 
Finally, one could ask whether cooperation would still spread starting with very few cooperators. Numerical simulations
 show that cooperation could indeed prevail even starting from as low as $1\%$ cooperators,
except on the far left border of the configuration space where cooperation is severely disadvantaged.\\
Compared with the level of cooperation observed in simulations in
static networks, we can say that results are consistently better for co-evolving networks.  For all values of
$q$ (Fig.~\ref{hd_coop}) there is significantly more cooperation than what was found in model and real social networks~\cite{luthi-pest-tom-physa07} where the same local replicator dynamics
was used but with the constraints imposed by the
invariant network structure.
A comparable high cooperation level has only been found in static
scale-free networks~\cite{santos-pach-05,santos-pach-06} which are not as realistic as a social network structures. \\
The above considerations are all the more interesting when one observes that the standard RD result is
that the only asymptotically stable state for the game is a polymorphic population in which there
is a fraction $\alpha$ of doves and a fraction $1-\alpha$ of hawks, with $\alpha$ depending on the actual numerical
payoff matrix values.
To see the positive  influence  of making and breaking ties we can compare our results with what is prescribed by the standard RD solution. Referring to the payoff table~\ref{payoffs}, let's assume that the column player plays C with
probability $\alpha$ and D with probability $1-\alpha$. In this case, the expected payoffs of the row player
are:
$$
E_r[C] =\alpha R + (1-\alpha)S$$
and
$$E_r[D] =\alpha T + (1-\alpha)P
$$

\noindent The row player is indifferent to the choice of $\alpha$ when $E_r[C] = E_r[D]$. Solving for $\alpha$ gives:

\begin{equation}
 \alpha = \frac{P-S}{R-S-T+P}.
 \label{alpha}
\end{equation}

Since the game is symmetric, the result for the column player is the same and $(\alpha C, (1-\alpha) D)$
is a NE in mixed strategies.
We have numerically solved the equation for all the sampled points in the game's parameter space. Let us now use the following payoff values 
 in order to bring them within the explored game space (remember that NEs
are invariant w.r.t. such an affine transformation):

\begin{table}[hbt]
%\vspace{-0.2cm}
\begin{center}
{\normalsize
\begin{tabular}{c|cc}
 & C & D\\
\hline
C & ($1,1$) & ($2/3,4/3$)\\
D & ($4/3,2/3$) & ($0,0$)
\end{tabular}
}
\end{center}
%\vspace{-0.5cm}
\end{table}

Substituting in equation~\ref{alpha} gives $\alpha=2/3$, i.e. the dynamically stable polymorphic population should be
composed by about $2/3$ cooperators and $1/3$ defectors. Now, if one looks at Fig.~\ref{hd_coop} at the
points where $S=2/3$ and $T=4/3$, one can see that the point, and the region around it, is
one of full cooperation instead.
Even within the limits of the approximations caused by the finite population size and the local dynamics,  the non-homogeneous graph structure and an increased level of tie rewiring has allowed cooperation to be greatly enhanced with respect to the theoretical predictions of standard RD.

\subsection{Evolution of Agents' Satisfaction}

\begin{figure} [!ht]
\begin{center}
%	\vspace*{-0.2cm}
\includegraphics[width=10cm] {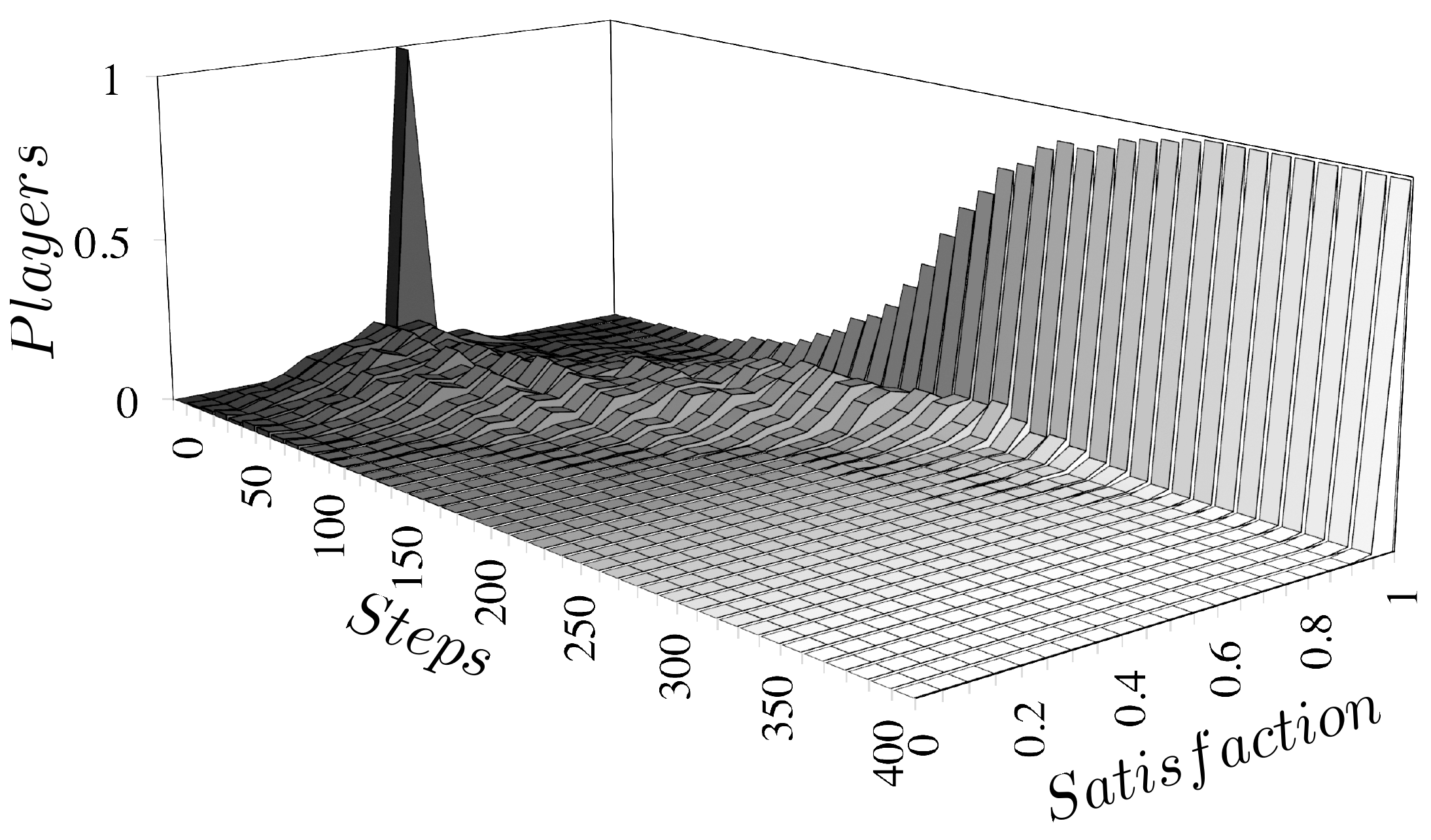} \protect \\	
%	\vspace*{-0.2cm}
\caption{Fraction of agents having a given satisfaction level as a function of evolution time.\label{sat}}
\end{center}
\end{figure}

According to the model, unsatisfied agents are more likely to try to cut links in an attempt to improve
their satisfaction level, which could be simply described as an average value of the strengths of their
links with neighbors. Satisfaction should thus tend to increase during evolution. In effect, this is what happens,
as can be seen in Fig.~\ref{sat}. The figure refers to a particular run that ends in all agents cooperating,
but it is absolutely typical.
One can remark the ``spike'' at time $0$. This is clearly due to the fact that all links are initialized
with a weight of $0.5$. As the simulation advances, the satisfaction increases steadily and for the case
of the figure, in which all agents cooperate at the end, it reaches its maximum value of $1$ for almost all
players.

\subsection{Stability of Cooperation}
\label{stability}

Evolutionary game theory provides a dynamical view of conflicting decision-making in populations.
Therefore, it is important to assess the \textit{stability} of the equilibrium configurations.
This is even more important in the case of numerical simulation where the steady-state
finite population configurations are not really equilibria in the mathematical sense. In other words,
one has to be reasonably confident that the steady-states are not easily destabilized
by perturbations. To this end, we have performed a numerical study of the robustness of
final cooperators' configurations by introducing a variable amount of random noise into
the system. A strategy is said to be \textit{evolutionarily stable} when it cannot be invaded
by a small amount of players using another strategy~\cite{hofb-sigm-book-98}. We have chosen 
to switch the strategy of
an increasing number of highly connected cooperators to defection, and to observe whether
the perturbation propagates in the population, leading to total defection, or if it stays 
localized and disappears after a transient time.
\begin{figure} [!ht]
\begin{center}
%	\vspace*{-0.2cm}
\includegraphics[width=14cm,height=4cm] {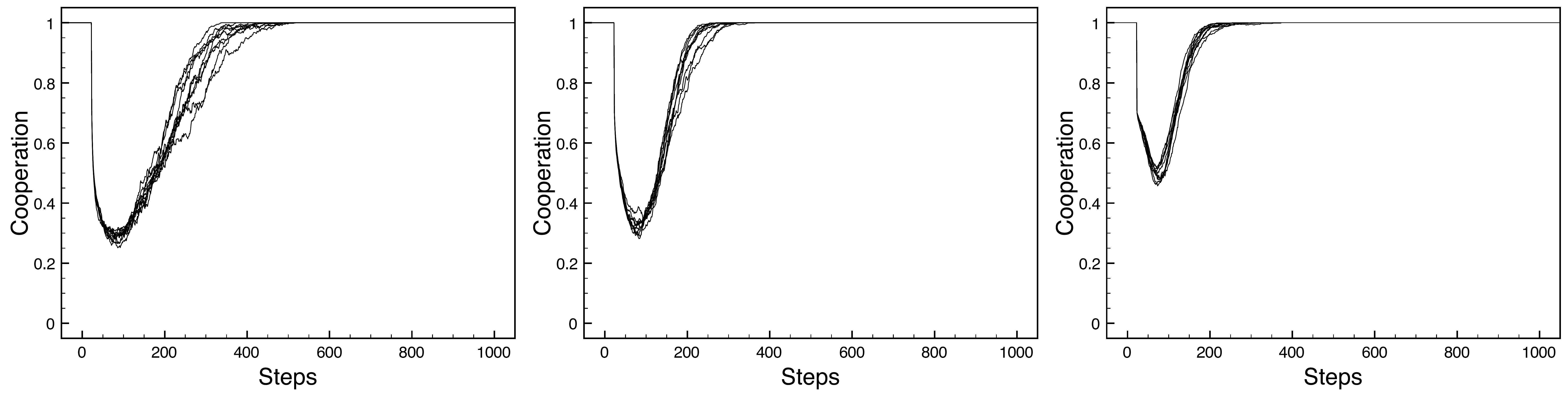} \protect \\	
%	\vspace*{-0.2cm}
\caption{Cooperation percentage as a function of simulated time when the strategy of the $30\%$ most connected nodes is switched
from cooperation to defection. $T=1.6$, $S=0.4$ and, from left to right, $q=0.2, 0.5, 0.8$.\label{hd_noise_16}}
\end{center}
\end{figure}
\begin{figure} [!ht]
\begin{center}
%	\vspace*{-0.2cm}
\includegraphics[width=14cm,height=4cm] {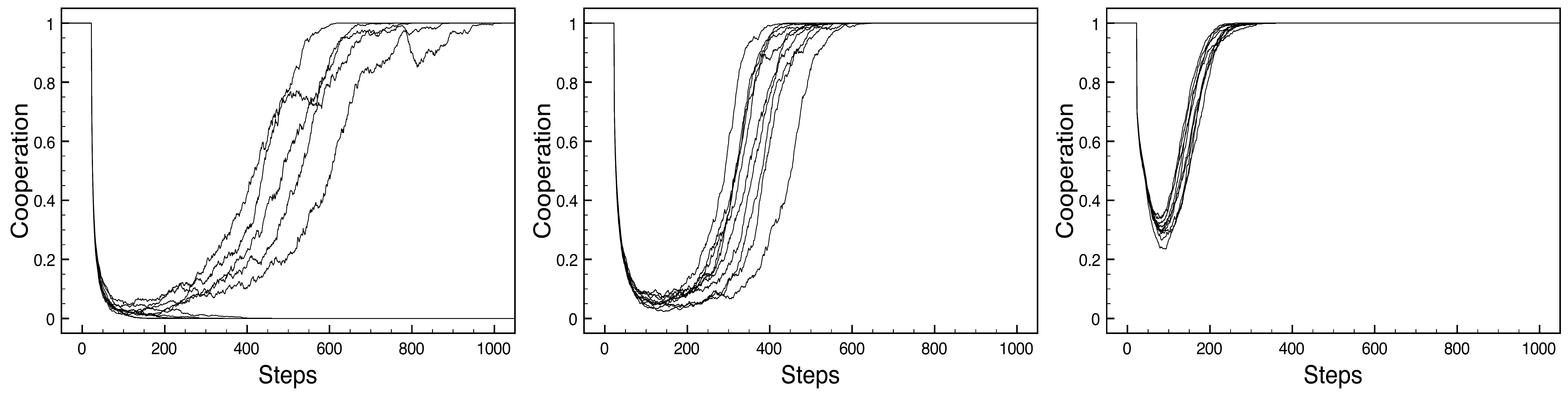} \protect \\	
%	\vspace*{-0.2cm}
\caption{Cooperation percentage when the strategy of the $30\%$ most connected nodes is switched
from cooperation to defection. $T=1.9$, $S=0.1$ and, from left to right, $q=0.2, 0.5, 0.8$.\label{hd_noise_19}}
\end{center}
\end{figure}
Figs.~\ref{hd_noise_16} and~\ref{hd_noise_19} show how the system recovers when the most highly
connected $30$\% of the cooperators are suddenly and simultaneously switched to defection. In Fig.
\ref{hd_noise_16} the value chosen in the game's configuration space is $T=1.6$ and $S=0.4$. This
point lies approximately on the diagonal in Fig.~\ref{hd_coop} and corresponds to an all-cooperate
situation. As one can see, after the perturbation is applied, there is a sizable loss of cooperation but,
after a while, the system recovers full cooperation in all cases (only $10$ curves are shown in each figure
for clarity, but the phenomenon is qualitatively identical in all the $50$ independent runs tried). From left to right,
three values of $q=0.2,0.5,0.8$ are used. It is seen that, as the rewiring frequency $q$ increases, recovering from the
perturbation becomes easier as defection has less time to spread around before cooperators are
able to dismiss links toward defectors. Switching the strategy of the $30$ \% most connected nodes
is rather extreme since they include most cooperator clusters but, nonetheless, cooperation is rather stable in the whole cooperating region.  In Fig.~\ref{hd_noise_19} we have done the same this time with
$T=1.9$ and $S=0,1$. This point is in a frontier region in which defection may often prevail, at least
for low $q$ (see Fig.~\ref{hd_coop}) and thus it represents one of the hardest cases for
cooperation to remain stable. Nevertheless, except in the leftmost case ($q=0.2$) where half of the runs
permanently switch to all-defect, in all the other cases the population is seen to recover after
cooperation has fallen down to less than $10\%$. Note that the opposite case is
also possible in this region that is, in a full defect situation, switching of $30\%$ highly connected defectors
to cooperation can lead the system to one of full cooperation.
In conclusion, the above numerical experiments have empirically shown that cooperation is
extremely stable after cooperator networks have emerged. Although we are using here an artificial
society of agents, this can hopefully be seen as an encouraging result for cooperation in real societies.

\subsection{Structure of the Emerging Networks}
\label{topo}

In this section we present a statistical analysis of the global and local properties of the networks that
emerge when the pseudo-equilibrium states of the dynamics are attained. Note that in the following sections
the graph we refer to is the unoriented, unweighted one that we called $G^{'}$ in Sect.~\ref{net}. In other words, for
the structural properties of interest, we only take into account the fact that two agents interact and not
the weights of their directed interactions.

\subsubsection{Small-World Nature}
Small-world networks are characterized by a small mean path length and by a high clustering coefficient~\cite{watts-strogatz-98}. Our graphs start random, and  thus have short path lengths by construction since 
their mean path length $\bar l= O(log N)$ scales logarithmically with the number of vertices $N$~\cite{newman-03}.
It is interesting to notice that they maintain short diameters at equilibrium too, after rewiring has taken
place. We took the average $\bar L = \sum_{k=1}^{660} \bar l$ of the mean path length of $660$ evolved graphs, which represent ten graphs for each $T,S$ pair. This average
 is 3.18, which is of the order of $log(1000)$, while its initial random graph average value is $3.25$. This fact, together with the remarkable increase of
the clustering coefficients with respect to the random graph (see below), shows that the evolved networks have the small-world property.
Of course, this behavior was expected, since the rewiring mechanism favors close partners in the network and thus
tends to increase the clustering and to shorten the distances.

\subsubsection{Average Degree} In contrast to other models~\cite{zimmer-pre-05,santos-plos-06}, the mean degree $\bar k$ can
vary during the course of the simulation. We found that $\bar k$ increases only slightly and tends to stabilize around $\bar k=11$. This is in qualitative agreement with observations made on real dynamical
social networks~\cite{kossi-watts-06,barab-collab-02,tom-leslie-evol-net-07} with the only difference that the network does not grow in our model.

\begin{figure} [!ht]
\begin{center}
%	\vspace*{-0.2cm}
\includegraphics[width=14.5cm] {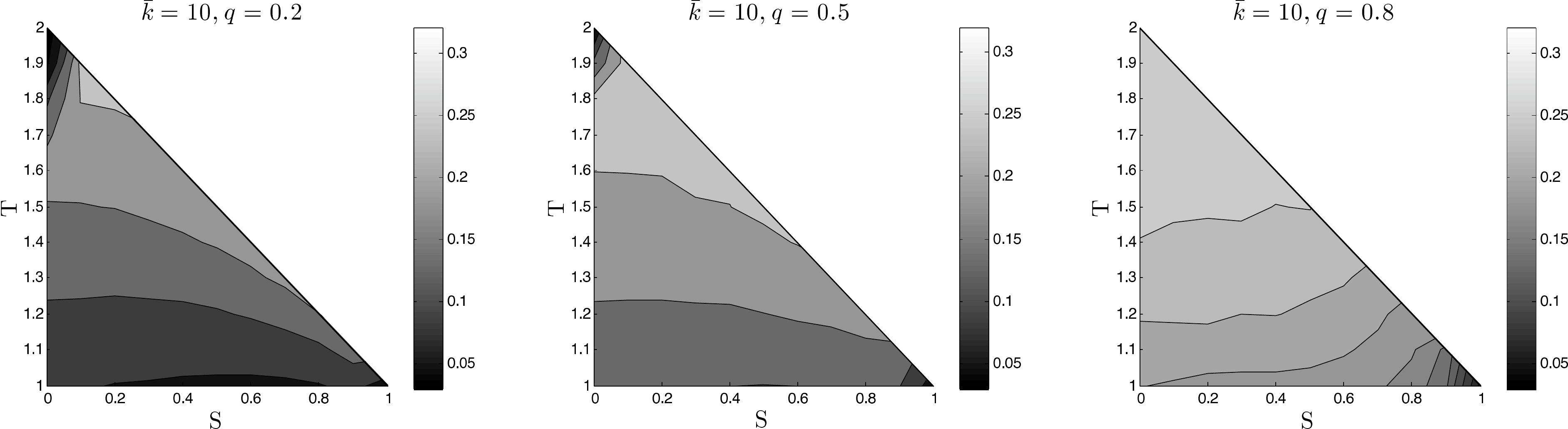} \protect \\	
%	\vspace*{-0.2cm}
\caption{Average values of the clustering coefficient over $50$ runs for three
values of $q$. \label{hd_clust}}
\end{center}
\end{figure}

\subsubsection{Clustering Coefficients}
The  clustering coefficient $\cal C$ of a graph has been defined in the Introduction section.
Random graphs are locally homogeneous in the average and for them $\cal C$ is simply equal to
the probability of having an edge between any pair of nodes independently. In contrast, real networks have
local structures and thus higher values of $\cal C$. Fig.~\ref{hd_clust} gives the average clustering coefficient
$\bar {\cal C}=\frac{1}{50} \sum_{i=1}^{50} {\cal C}$ for
each sampled point in the Hawks-Doves configuration space, where $50$ is the number of network realizations
used for each simulation. 
The networks self-organize through dismissal of partners and choice of new ones and they
acquire local structure, since the clustering coefficients 
are higher than that of a random graph
with the same number of edges and nodes, which is $\bar k/N=10/1000=0.01$.
The clustering tends to increase with $q$ (i.e.~from left to right in Fig.~\ref{hd_clust}). It is clear that
the increase in clustering and the formation of cliques is  due to the fact that, when dismissing
an unprofitable relation and searching for a new one, individuals that are relationally at a short distance
are statistically favored. But this has a close correspondence in the way in which new
acquaintances are made in society: they are not random, rather people often get to interact with each
other through common acquaintances, or ``friends of friends'' and so on.

\subsubsection{Degree Distributions}

\begin{figure} [!ht]
\begin{center}
%	\vspace*{-0.2cm}
\includegraphics[width=8cm] {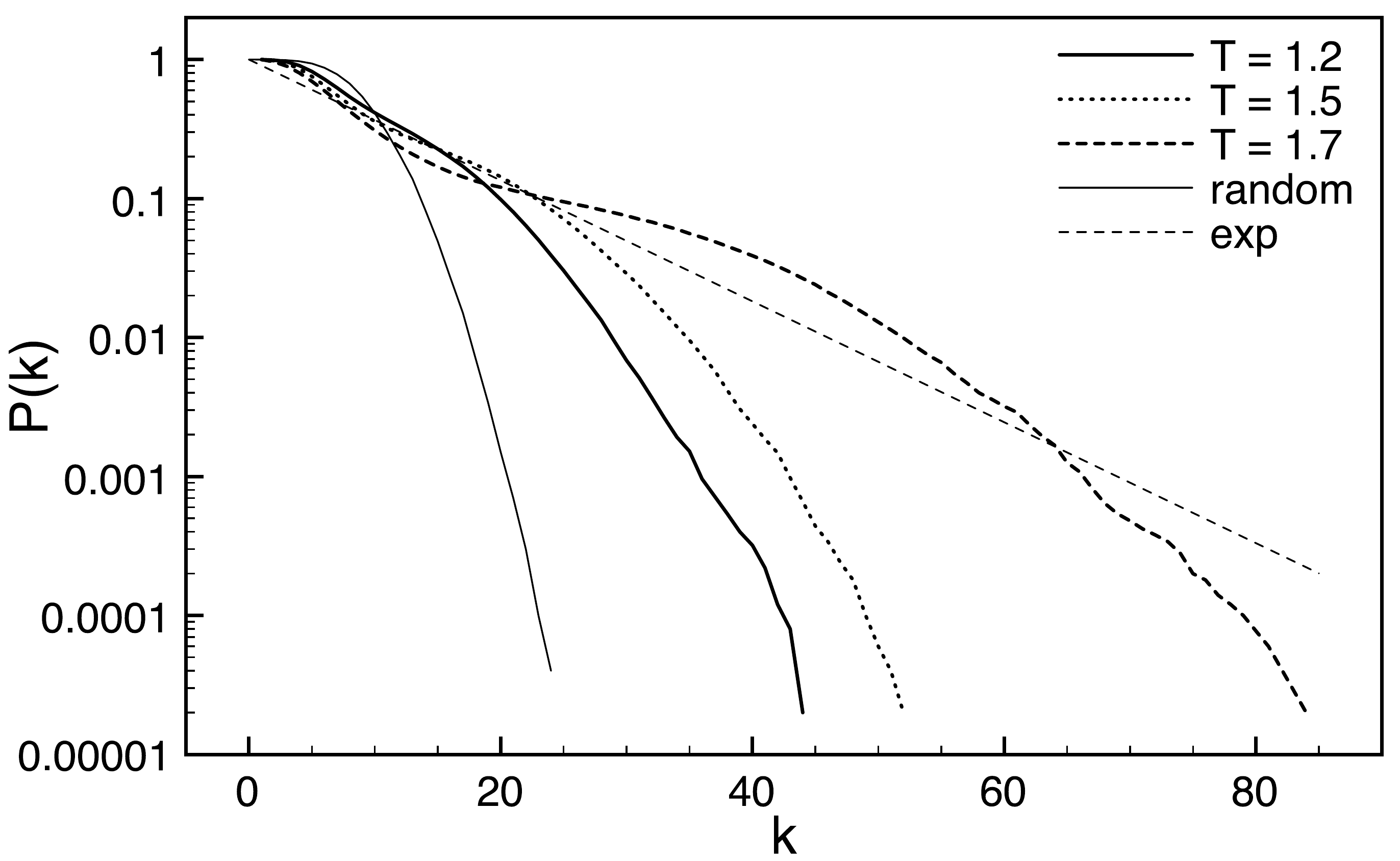} \protect \\	
%	\vspace*{-0.2cm}
\caption{Empirical cumulative degree distribution functions for three different values of the
temptation $T$. A Poissonian and an exponential distribution are also plotted for
comparison. Distributions are discrete, the continuous lines are only a guide for
the eye. Lin-log scales.\label{hd_df}}
\end{center}
\end{figure}

\begin{figure} [!ht]
\begin{center}
%	\vspace*{-0.2cm}
\includegraphics[width=8cm] {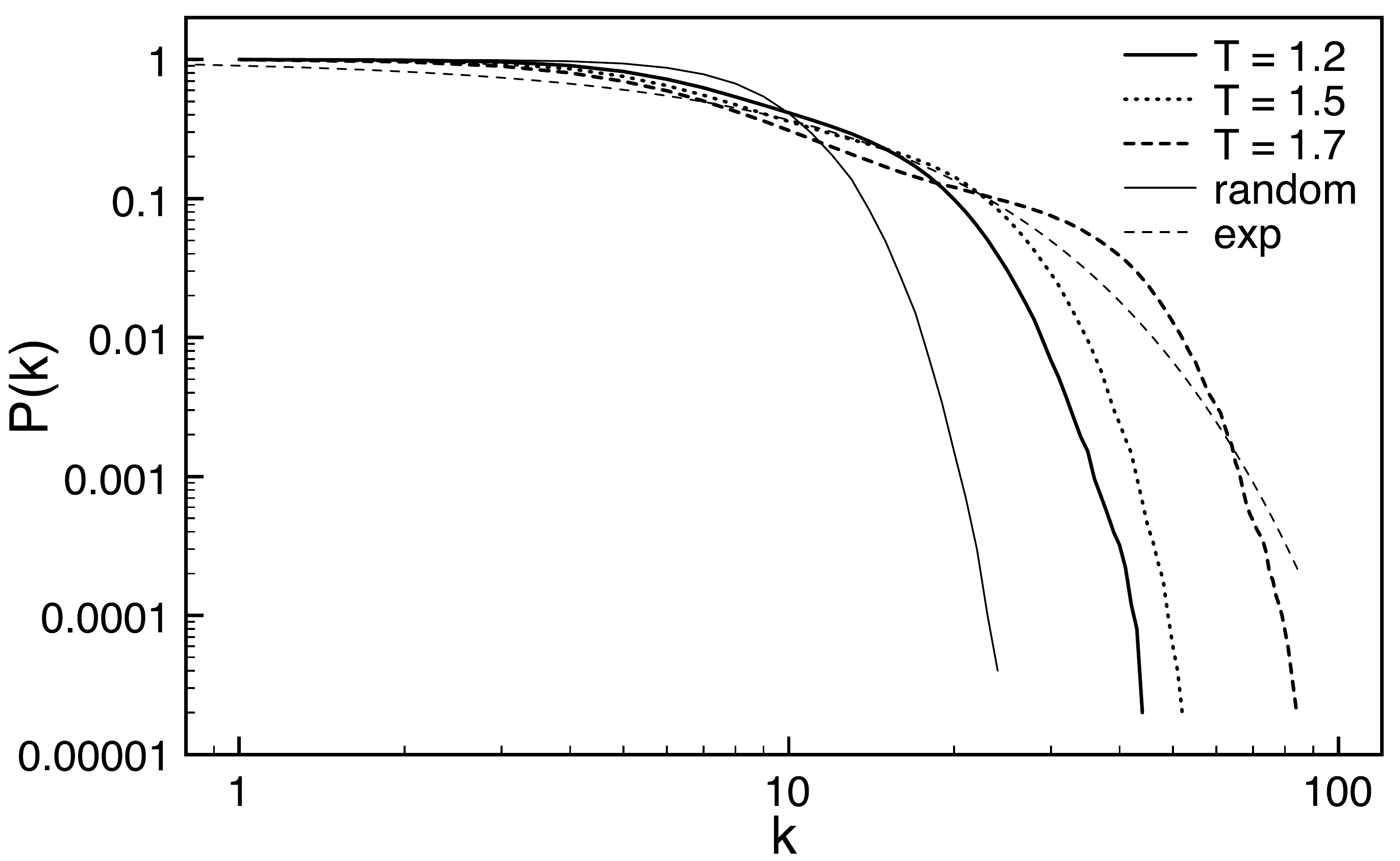} \protect \\	
%	\vspace*{-0.2cm}
\caption{Empirical cumulative degree distribution functions for three different values
of the parameter $T$.  Log-log scales.\label{cddf-ll}}
\end{center}
\end{figure}

The \textit{degree distribution function} (DDF) $p(k)$  of a graph represents the probability that a randomly
chosen node has degree $k$. Random graphs are characterized by DDF of Poissonian
form  $p(k) = {\bar k}^k e^{-\bar k}/k!$, while social and technological real networks often show long tails to the right, i.e. there are nodes
that have an unusually large number of neighbors~\cite{newman-03}. In some extreme cases
the DDF has a power-law form $p(k) \propto k^{-\gamma}$; the 
tail is particularly extended and there is no characteristic degree. The \textit{cumulative
degree distribution function} (CDDF) is just the probability that the degree is greater than or equal to $k$
and has the advantage of being less noisy for high degrees.
 Fig.~\ref{hd_df} shows the CDDFs for the Hawks-Doves for three values of $T$, $S=0.2$, and $q=0.5$ with
 a logarithmic scale on the y-axis. A Poisson and an exponential distribution are also included for comparison. The Poisson curve actually represents the
 initial degree distribution of the (random) population graph.  The distributions at equilibrium are
 far from the Poissonian that would apply if the networks would remain essentially random. However, they
 are also far from the power-law type, which would appear as a straight line in the log-log plot of
 Fig~\ref{cddf-ll}. 
 Although a reasonable fit with a single law appears to be difficult, these empirical distributions are closer
 to exponentials, in particular the curve for $T=1.7$, for which such a fit has been drawn.
 It can be observed that the
 distribution is broader the higher $T$ (The higher $T$, the more agents gain by defecting). In fact, although cooperation is attained nearly
 everywhere in the game's configuration space, higher values of the temptation $T$ mean that
 agents have to rewire their links more extensively, which results in a higher number of neighbors
 for some players, and thus it leads to a longer tail in the CDDF.
 \begin{figure} [!ht]
\begin{center}
%	\vspace*{-0.2cm}
\includegraphics[width=8cm] {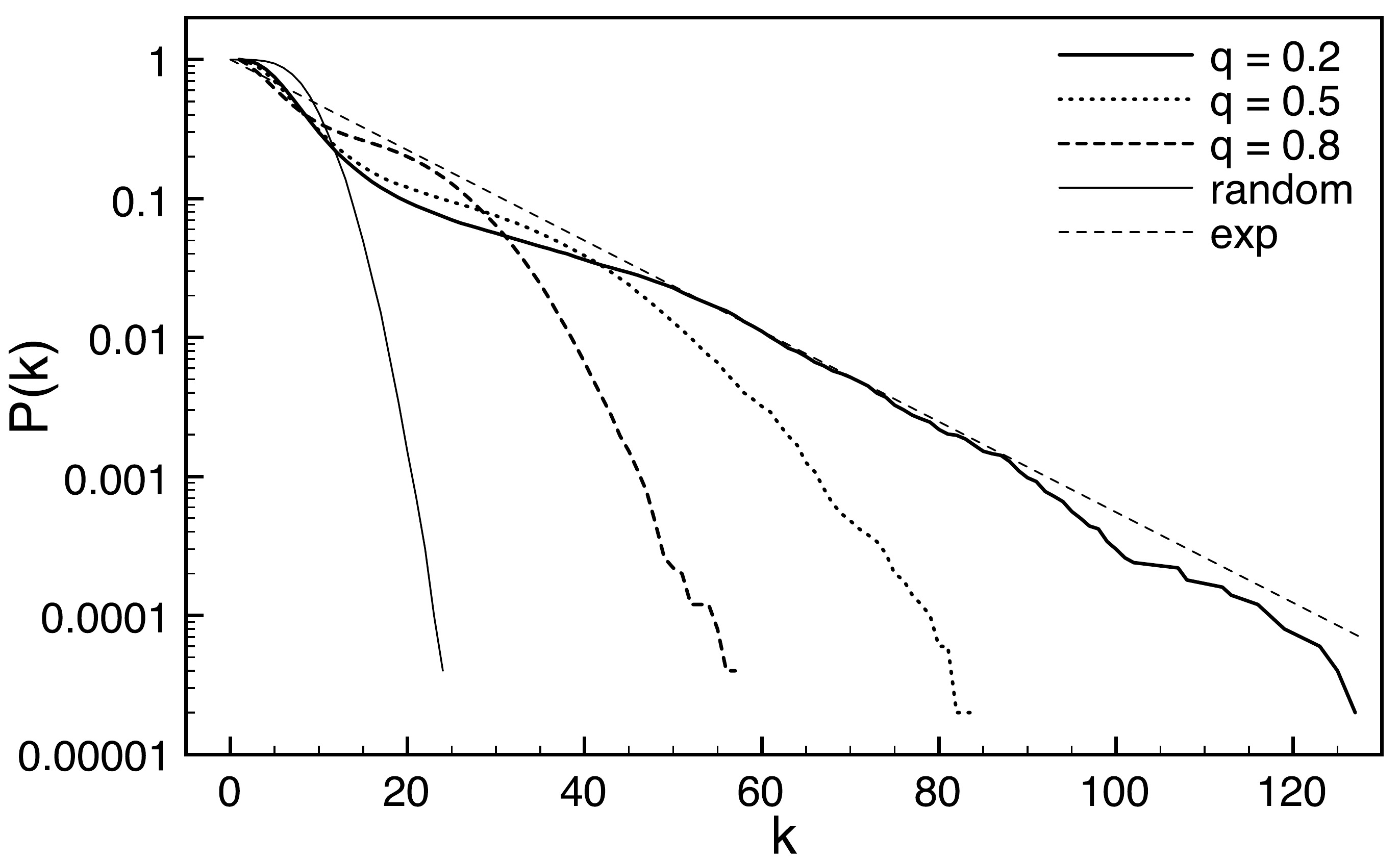} \protect \\	
%	\vspace*{-0.2cm}
\caption{Empirical cumulative degree distribution functions for three different values of the
temptation $q$.  Lin-log scales.\label{hd-cddf-q}}
\end{center}
\end{figure}
The influence of the $q$ parameter on the shape of the degree
distribution functions is shown in Fig.~\ref{hd-cddf-q} where average curves for three values of $q$, $T=1.7$, and $S=0.2$, are
reported. For high $q$, the cooperating steady-state is reached faster, which gives the network less time to  rearrange its links. For lower values of $q$ the distributions become broader, despite the fact that
rewiring occurs less often, because cooperation in this region is harder to attain and more simulation time
is needed. In conclusion, emerging network structures at steady states have DDFs that are similar to those found
in actual social networks~\cite{newman-03,am-scala-etc-2000,newman-collab-01-1,jordano03,TLGL-GPEM-07}, with tails that are fatter the higher the temptation $T$ and the lower $q$. Topologies closer to scale-free would probably be obtained if
the model allowed for growth, since preferential attachment is already present to some extent due to the
nature of the rewiring process~\cite{poncela08}.

\subsubsection{Degree Correlations}
  
Besides the degree distribution function of a network, it is also sometimes useful to
investigate the empirical joint degree-degree distribution of neighboring vertices. However,
it is difficult to obtain reliable statistics because the data set is usually too small (if a network
has $L$ edges, with $L \ll N^2$ where $N$ is the number of vertices for the usually
relatively sparse networks we deal with, one then has only $L$ pairs of data to work with).
Approximate statistics can readily be obtained by using the average degree of the
nearest neighbors of a vertex $i$ as a function of the degree of this vertex, $\bar k_{V_i}(k_i)$~\cite{PS-VAZ-VESP}.
 \begin{figure} [!ht]
\begin{center}
%	\vspace*{-0.2cm}
\includegraphics[width=7.5cm] {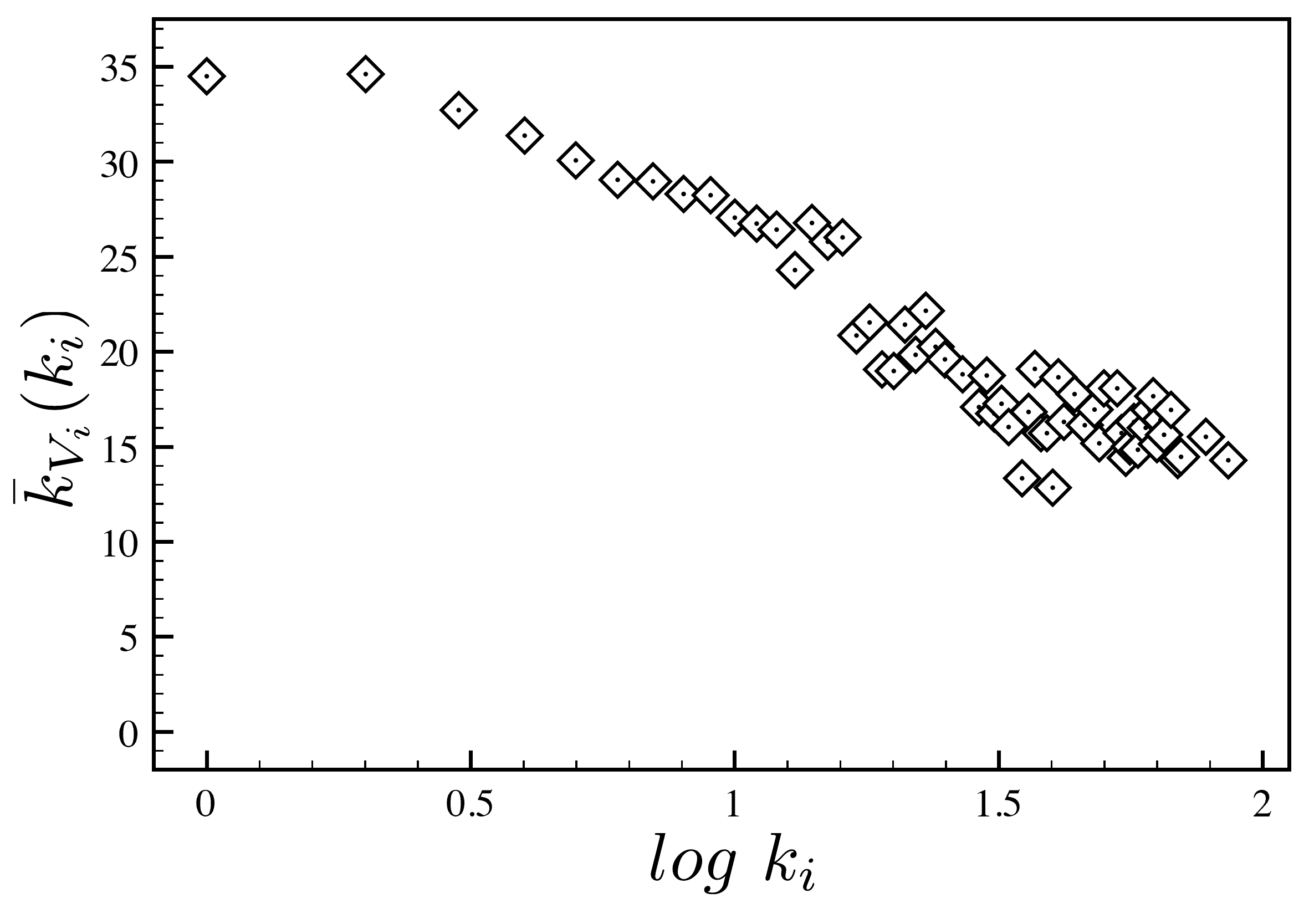} \protect \\	
%	\vspace*{-0.2cm}
\caption{Average degree of the direct neighbors of a vertex Vs. the vertex degree. The relation is
disassortative. Log-lin scales.\label{ddc}}
\end{center}
\end{figure}
From Fig.~\ref{ddc} one can see that the correlation is slightly negative, or disassortative. This is at odds with
what is reported about real social networks, in which usually this correlation is positive instead,
i.e. high-degree nodes tend to connect to high-degree nodes and vice-versa~\cite{newman-03}. However,
real social networks establish and grow because of common interests, collaboration work, friendship
and so on. Here this is not the case, since the network is not a growing one, and the game played by the 
agents is antagonistic and causes segregation of highly connected cooperators into clusters in which they  
are surrounded by less highly connected fellows. This will be seen more pictorially in the following section.

\subsection{Cooperator Clusters}
\label{comm}

\begin{figure} [!ht]
\begin{center}
%	\vspace*{-0.2cm}
	
\includegraphics[width=8cm] {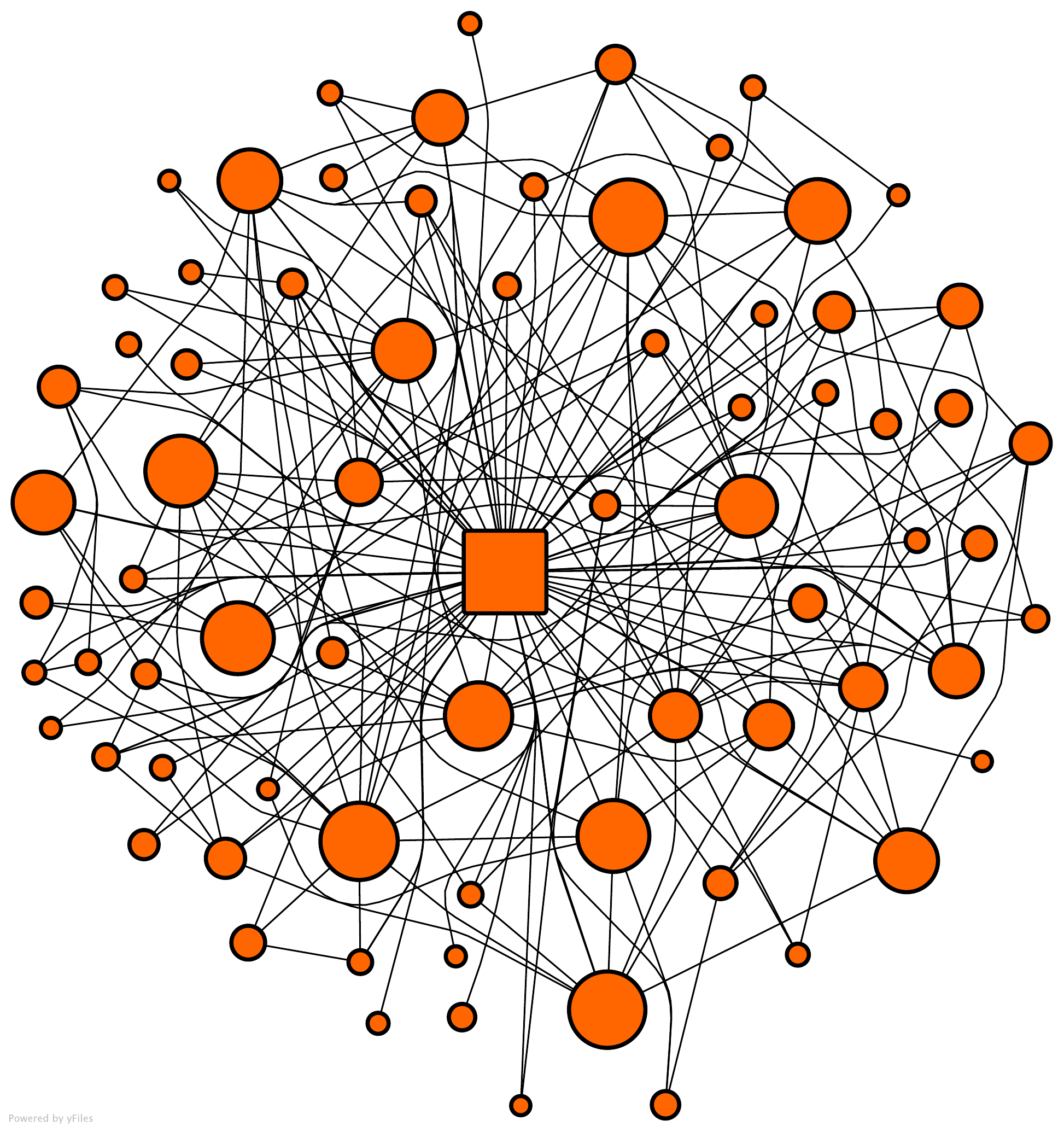} \protect \\	
%	\vspace*{-0.2cm}
\caption{A typical cooperator cluster. Links to the rest of the network have been suppressed
for clarity. The size of a node is proportional to its connectivity in the whole graph. The most connected central cooperator is shown as a square.\label{hd_cluster}}
\end{center}
\end{figure}

From the results of the previous sections, it appears that a much higher amount of cooperation
than what is predicted by the standard theory for mixing populations can be reached when
ties can be broken and rewired. We have seen that this dynamics causes the graph to
acquire local structure, and thus to loose its initial randomness. In other words,
the network self-organizes in order to allow players to cooperate as much as possible.
At the microscopic, i.e.~agent level, this happens
 through the formation of clusters of
players using the same strategy. Fig.~\ref{hd_cluster} shows one typical cooperator cluster. 
In the figure one can clearly see that the central cooperator
is a highly connected node and there are many links also between the other neighbors. Such
tightly packed structures have emerged to protect cooperators from defectors that, at earlier times, were
trying to link to cooperators to exploit them. These observations help understand why the degree  distributions are long-tailed (see previous section), and also the higher values of the clustering coefficient.\\
Further studies of the emerging networks would imply investigating the communities and the way in which
strategies are distributed in them. There
are many ways to reveal the modular structure of networks~\cite{Arenas05} but we leave this study for further work.

\section{Conclusions}
\label{concl}

In this paper we have introduced a new dynamical population structure for agents playing
a series of two-person Hawks and Doves game. The most novel feature of the model is the
adoption of a variable strength of the bi-directional social ties between pairs of players. These strengths
change dynamically and independently as a function of the relative satisfaction of the two end points
when playing with their immediate neighbors in the network.
A player may wish to break a tie to a neighbor and the probability of cutting the link is higher
the weaker the  directed link strength is. The ensemble of weighted links implicitly represents
a kind of memory of past encounters although, technically speaking, the game is not iterated.
While in previous work the rewiring parameters where ad hoc, unspecified probabilities,
we have made an effort to relate them to the agent's propensity to gauge the perceived
quality of a relationship during time. \\
The model takes into account recent knowledge coming from the analysis of the structure and
of the evolution of social networks and, as such, should be a better approximation of real
social conflicting situations than static graphs such as regular grids. In particular, new links are not created
at random but rather taking into account the ``trust'' a player may have on her relationally close social
environment as reflected by the current strengths of its links. This, of course, is at the origin of the de-randomization and self-organization of the network,
with the formation of stable clusters of cooperators. The main result concerning
the nature of the pseudo-equilibrium states of the dynamics is that cooperation is greatly
enhanced in such a dynamical artificial society and, furthermore, it is quite robust with respect to
large strategy perturbations. Although our model is but a simplified and incomplete
representation of social reality, this is encouraging, as the Hawks-Doves game
is a paradigm for a number of social and political situations in which aggressivity plays an
important role. The standard result is that bold behavior does not disappear at evolutionary
equilibrium. However, we have seen here that a certain amount of plasticity of the networked
society allows for full cooperation to be consistently attained. Although the model is an extremely abstract one,
it shows that there is place for peaceful resolution of conflict. In future work we would like to investigate
other stochastic strategy evolution models based on more refined forms of learning than
simple imitation and study the global modular structure of the equilibrium networks.\\

{\bf Acknowledgements.}
This work is funded by  the Swiss National Science Foundation under grant number 200020-119719.
We gratefully acknowledge this financial support.

\bibliographystyle{elsart-num}

%\bibliography{jeux}

\end{document}